\documentclass[pre,twocolumn]{revtex4-1}
\usepackage{amsmath, color}
\usepackage{indentfirst}
\usepackage{amssymb}
\usepackage{graphicx}
\usepackage{epsfig}
\usepackage{float}

\newcommand{\C}[1]{{\mathcal{#1}}}

\begin{document}
\title{Variable-amplitude oscillatory shear response of amorphous materials}
\author{Nathan Perchikov and Eran Bouchbinder}
\affiliation{Department of Chemical Physics, Weizmann Institute of Science, Rehovot 76100, Israel}

\begin{abstract}
Variable-amplitude oscillatory shear tests are emerging as powerful tools to investigate and quantify the nonlinear rheology of amorphous solids, complex fluids and biological materials. Quite a few recent experimental and atomistic simulation studies demonstrated that at low shear amplitudes, an amorphous solid settles into an amplitude- and initial conditions-dependent dissipative limit cycle, in which back-and-forth localized particle rearrangements periodically bring the system to the same state. At sufficiently large shear amplitudes, the amorphous system loses memory of the initial conditions, exhibits chaotic particle motions accompanied by diffusive behavior and settles into a stochastic steady-state. The two regimes are separated by a transition amplitude, possibly characterized by some critical-like features. Here we argue that these observations support some of the physical assumptions embodied in the nonequilibrium thermodynamic, internal-variables based, Shear-Transformation-Zone model of amorphous visco-plasticity; most notably that ``flow defects'' in amorphous solids are characterized by internal states between which they can make transitions, and that structural evolution is driven by dissipation associated with plastic deformation. We present a rather extensive theoretical analysis of the thermodynamic Shear-Transformation-Zone model for a variable-amplitude oscillatory shear protocol, highlighting its success in accounting for various experimental and simulational observations, as well as its limitations. Our results offer a continuum-level theoretical framework for interpreting the variable-amplitude oscillatory shear response of amorphous solids and may promote additional developments.
\end{abstract}

\maketitle

\section{Background and motivation}
\label{sec:intro}

The application of oscillatory shear deformation of the form $\gamma(t)\!=\!\gamma_0\sin(\omega\,t)$ -- where $\gamma(t)$ is the time-dependent shear strain, $t$ is time, $\gamma_0$ is the shear amplitude and $\omega$ is the oscillations frequency -- offers an important protocol to probe the rheological properties of a broad range of physical systems, including amorphous solids, complex fluids and biological materials. The most well-developed and well-documented rheological test in this context focusses on the linear viscoelastic response, where the amplitude $\gamma_0$ is very small and the frequency $\omega$ is systematically varied. In this case, the steady-state linear stress response is fully characterized by a single frequency-dependent complex function $G^*(\omega)\!=\!G'(\omega)+i G''(\omega)$, where the storage (shear) modulus $G'(\omega)$ quantifies the linear elastic response and the loss modulus $G''(\omega)$ quantifies the linear viscous response.

A complementary protocol which sheds light on nonlinear material rheology, and the transition between linear and nonlinear rheologies, is obtained by fixing the frequency $\omega$ and systematically varying the amplitude $\gamma_0$, within and well beyond the linear response regime. Such variable-amplitude oscillatory shear tests applied to amorphous solids have been the focus of quite a few recent simulational and experimental studies \cite{Hebraud1997, Petekidis2002, Lacks2004, Lundberg2008, Slotterback2012, Ren2013, Priezjev2013, Keim2013a, Fiocco2013, Regev2013, Keim2014, Priezjev2014, Nagamanasa2014}. These include experiments on disordered emulsions \cite{Hebraud1997}, colloidal glasses \cite{Petekidis2002, Nagamanasa2014}, amorphous foams \cite{Lundberg2008}, granular packings \cite{Slotterback2012, Ren2013}, and jammed interfacial materials \cite{Keim2013a, Keim2014}; and simulations of computer-generated binary mixture glasses at zero temperature \cite{Lacks2004, Fiocco2013, Regev2013} and finite temperatures \cite{Priezjev2013}, a bubble model of two-dimensional foams \cite{Lundberg2008} and a bead-spring model of low molecular-weight polymer glasses \cite{Priezjev2014}. The generic physical picture of the variable-amplitude oscillatory shear response of amorphous solids emerging from these studies -- to be described in detail below -- is not yet fully understood and accounted for in the framework of a continuum theory.

Our goal in this paper is to theoretically study this variable-amplitude oscillatory shear protocol in the framework of the ``plain vanilla'' nonequilibrium thermodynamic, internal-variables based, Shear-Transformation-Zone (STZ) model of amorphous visco-plasticity. This model, to be summarized below, has been shown to capture several salient features of the phenomenology of driven amorphous systems \cite{98FL, JSL04, BLP-07-I, BLP-07-II, Manning, 08JSL, BLIII-09, LinearResponse_PRL_2011, LinearResponse_PRE_2011, 11FL, Bouchbinder2013}. Yet, it has not -- up until now -- been analyzed under nonlinear oscillatory shear protocols. Such an analysis both challenges the underlying assumptions of the model and has the potential to shed light on the basic physics of driven amorphous solids, an important open problem in condensed-matter and materials physics, with far reaching implications to basic science and technology.

Oscillatory shear protocols can in fact involve systematically varying both the amplitude $\gamma_0$ and frequency $\omega$ over a broad range, providing a comprehensive characterization of the rheological properties of materials. Indeed, large amplitude oscillatory shear (LAOS) tests offer powerful tools to investigate the nonlinear rheology of materials, see for example \cite{Hyun2011, Koumakis2013, Ewoldt2013}. Theoretical analyses based on the Mode Coupling Theory (MCT) for dense colloidal suspensions \cite{Brader2010} and the Soft Glassy Rheology model (SGR) \cite{Sollich1998} have been shown to capture some aspects of the steady-state nonlinear rheology of soft amorphous solids. A Shear-Transformation-Zone (STZ) analysis, which is based on coarse-grained structural internal-variables within a nonequilibrium thermodynamic framework, and which also addresses non-steady-state dynamics, can complement these recent studies in interesting ways. It may also promote additional developments toward a predictive macroscopic theory of the mechanics of amorphous solids.

The structure of the paper is as follows. In Sect. \ref{sec:phenomenology} the phenomenology of the variable-amplitude oscillatory shear response of amorphous solids is described. The thermodynamic STZ model is briefly reviewed in Sect. \ref{sec:STZmodel} and theoretically analyzed for a variable-amplitude oscillatory shear protocol in Sect. \ref{sec:theory}. The relation of the theoretical results to recent experimental and simulational observations is discussed in Sect. \ref{sec:exp} and some concluding remarks are offered in Sect. \ref{sec:sum}.

\section{The phenomenology of the variable-amplitude oscillatory shear response of amorphous solids}
\label{sec:phenomenology}

We start by summarizing the major observations made in very recent simulational and experimental studies on the variable-amplitude oscillatory shear response of amorphous solids \cite{Lundberg2008, Slotterback2012, Ren2013, Priezjev2013, Keim2013a, Fiocco2013, Regev2013, Keim2014, Priezjev2014, Nagamanasa2014}. In these studies, as explained above, amorphous solids (e.g computer-generated glasses, foams, granular materials, colloidal glasses) are subjected to oscillatory shear deformation of the form $\gamma(t)\!=\!\gamma_0\sin(\omega t)$, where the frequency $\omega$ is fixed and the amplitude $\gamma_0$ is systematically varied. The main phenomenology emerging from these studies is the following:\\

$\bullet$ At small amplitudes $\gamma_0$, the amorphous system settles into a $\gamma_0$-dependent dissipative limit cycle in which structural rearrangements repeatedly take place and the system returns exactly to the same state after an integer multiple of the loading period. Particles motion in this regime is non-diffusive. This behavior has been established in strictly athermal (zero temperature) systems and is decorated by thermal effects at finite temperatures.\\

$\bullet$ As long as $\gamma_0$ is smaller than some critical amplitude $\gamma_c$, $\gamma_0\!<\!\gamma_c$, the $\gamma_0$-dependent limit cycle is reached on a timescale that increases with $\gamma_0$. Some studies suggested a (possibly divergent) power-law increase of this timescale as $\gamma_0\!\to\!\gamma_c^{-}$.\\

$\bullet$ For $\gamma_0\!<\!\gamma_c$, the system retains memory of its initial state. That is, the states visited by the system remain correlated to some degree with the initial state for long (or even indefinitely long) times.\\

$\bullet$ As $\gamma_0$ surpasses the critical amplitude $\gamma_c$, i.e. for $\gamma_0\!>\!\gamma_c$, the system's response changes qualitatively. In particular, the dissipation per loading cycle in the oscillatory steady-state, i.e. the area under the hysteresis curve, grows dramatically as $\gamma_c$ is surpassed. Particles motion in this regime is chaotic and diffusive.\\

$\bullet$ For $\gamma_0\!>\!\gamma_c$, the system loses memory of its initial state. In particular, the system approaches a stochastic stationary state independent of the initial condition. Whether or not this steady-state is also $\gamma_0$-independent is to be discussed below.\\

$\bullet$ For $\gamma_0\!>\!\gamma_c$ the time to reach a steady-state falls off as $\gamma_0$ increases above $\gamma_c$. Some studies on athermal systems suggested a (possibly divergent) power-law increase of this time as $\gamma_0\!\to\!\gamma_c^{+}$.\\

Similar phenomenology has been observed in shear driven colloidal suspensions of dilute (well below the colloidal glass transition) non-Brownian particles \cite{Pine2005, Corte2008}. Recent simulations of frictionless athermal disks \cite{Schreck2013}, performed at intermediate packing fractions between that of contact percolation and that of the onset of jamming, might suggest some interesting connections between the solid-like response and the dense fluid-like response. The present study, though, focusses on the solid response.

The generic phenomenology described above calls for theoretical understanding. In what follows we test whether, and to what extent, this variable-amplitude oscillatory shear response of amorphous solids can be predicted by the continuum-level, thermodynamic Shear-Transformation-Zone (STZ) model.

\section{A brief overview of the thermodynamic STZ model}
\label{sec:STZmodel}

The ultimate goal of a macroscopic theory of plasticity is to predict the plastic strain rate $\dot\gamma^{pl}$ (generally a tensor, though tensorial notation is omitted here and below) as a function of the stress, the temperature and a set of structural internal-variables (order parameters) that carry information about the history of the deforming material (both preparation and deformation history). Achieving this goal remains a long-standing challenge in condensed-matter physics, statistical physics and materials science. While many different approaches were developed in this context, e.g. \cite{SOLLICH-97, Demetriou2006, BRADER-CATES-08,BRADER-CATES-09}, we focus here on the STZ model.

The STZ model is a continuum level model based on coarse-grained internal-variables and formulated within a nonequilibrium thermodynamic framework. It is based on an intrinsically nonperturbative approach and as such is not rigorously derived from first-principles many-body physics. Consequently, it involves some degree of phenomenology, which is directly inspired by molecular observations and constrained by symmetry principles, the laws of thermodynamics, physical insight and agreement with experiments. The model has been applied to a variety of problems in the mechanics of amorphous solids and has been shown to successfully capture several salient features of this important class of materials \cite{11FL}.

A notable recent example is the prediction of an annealing-induced brittle-to-ductile transition in metallic glasses \cite{Rycroft2012}. There, the STZ model has been used to calculate the material resistance to crack propagation in the presence of a pre-existing notch, whose surface evolves in time in response to external loading (this is a basic nonlinear material property, the so-called ``notch fracture toughness''). We believe that such a continuum approach, formulated in terms of a small set of partial differential equations, is necessary if we are to solve complex boundary-value problems in the mechanics of amorphous solids.

The STZ model has been presented, discussed and reviewed in several papers in recent years \cite{BLP-07-I, BLIII-09, 11FL}. Here we briefly outline the basic physical picture underlying the model and its mathematical formulation:\\

$\bullet$ The degrees of freedom of an amorphous solid are approximately separated into two weakly coupled
subsystems -- the slow configurational degrees of
freedom, i.e., the inherent structures -- and the fast
kinetic-vibrational degrees of freedom. The latter are characterized by the temperature of the heat reservoir, while the former are characterized by a thermodynamic temperature (sometimes termed ``effective temperature'' or ``configurational temperature'') that might differ from the heat reservoir temperature. Within a nonequilibrium thermodynamic framework, the effective temperature evolves through a configurational heat equation \cite{BLI-09, BLII-09}.\\

$\bullet$ Plastic deformation in amorphous materials is mediated by sparsely distributed localized zones, which are significantly more susceptible to shear rearrangements than their surroundings, and which are composed of a small number of basic elements (e.g. atoms, molecules, grains, bubbles etc.). These localized regions are termed Shear-Transformation-Zones (STZ). A nonequilibrium thermodynamic analysis suggests that the density of STZ, one of the coarse-grained internal-variables of the model, is determined by the effective temperature through a Boltzmann-like factor.\\

$\bullet$ A basic distinction is made between the existence of an STZ and its internal states. During its lifetime, an STZ can make transitions between its internal states at a rate that depends on the local stress and the heat reservoir temperature. In a minimal model, an STZ features two internal states, each of which is aligned along one of the two principal stress directions (in 2D), where transitions between these states make unit contributions to the macroscopic plastic strain rate. The average difference between the STZ sub-populations aligned in the two directions (normalized relative to the total population) defines the second internal-variable of the model, which quantifies deformation-induced structural anisotropy. It is important to note that the STZ transitions are distinctly different from macroscopic yielding, which is a global collective phenomenon. This is especially important in the present context.\\

$\bullet$ STZ can be created and annihilated by thermal and mechanical noise; the latter is generated by the irreversible rearrangements themselves and in the continuum description is taken to be proportional to the energy dissipation rate.\\

The mathematical formulation of this physical picture leads to the following expression for the plastic strain rate \cite{BLP-07-I, 11FL}
\begin{eqnarray}
\label{Dpl}
&&\dot\gamma^{pl}(s,T,\chi,m)=\\
&&\epsilon_0 e^{-1/\chi} e^{-1/T}  \cosh{\left(\frac{\Omega s}{T}\right )} \left[\tanh{\left ( \frac{\Omega s}{T}\right )}-m\right]\nonumber \ ,
\end{eqnarray}
already expressed in a dimensionless form. Note that tensorial notation is suppressed as we specialize here for simple shear deformation. In the expression above time is measured in units of molecular time $\tau$ and $s$ is the shear stress measured in units of a stress scale $s_y$ (which is {\em a posteriori} identified with the dynamic yield stress, hence the subscript). $\chi$ is the effective temperature measured in units of $e_z/k_B$, where $e_z$ is an STZ formation energy and $k_B$ is Boltzmann's constant, and $T$ is ordinary temperature measured in units of $\Delta/k_B$, where $\Delta$ is an activation energy barrier for STZ transitions. $\Omega\!\equiv\!\bar{\Omega}\epsilon_0 s_y/\Delta$, where $\bar{\Omega}$ is a typical volume of an STZ and $\epsilon_0$ is the typical local irreversible shear strain accumulated during an STZ transition (for thermal activation to be valid, $\Omega|s|$ cannot be too close to unity, a condition satisfied throughout our calculations).

$m$ is a coarse-grained internal state variable (somewhat analogous to the magnetization in magnetic systems), which quantifies deformation-induced structural anisotropy. In terms of the STZ degrees of freedom, as mentioned above, $m$ is defined as a normalized difference between STZ orientated along the two principal directions of the stress tensor. In an isotropic system we have $m\!=\!0$, as all STZ orientations are equally probable. $m$ plays an essential role in the analysis presented below. The Boltzmann-like factor $e^{-1/\chi}$ represents the probability to observe an STZ. $e^{-1/T}  \cosh{\left ( \frac{\Omega s}{T}\right )}$ represents a stress-biased thermal activations process. We term this model the ``plain vanilla'' thermodynamic STZ model because it incorporates only a single activation energy barrier and not a distribution of barriers \cite{LinearResponse_PRL_2011,LinearResponse_PRE_2011}.

$\dot\gamma^{pl}$ depends on four dynamical variables. As we restrict ourselves in the present study to slow deformation, we assume that ordinary heat generated during the deformation is quickly removed to the heat reservoir and hence that the ordinary thermal temperature $T$ is fixed at a value determined by the reservoir. The orientational internal variable $m$ evolves according to \cite{BLP-07-I, 11FL}
\begin{eqnarray}
\label{dotm}
&&\dot{m} = \frac{2 \dot\gamma^{pl}}{\epsilon_0 e^{-1/\chi}} \left(1-m s\right) =\\
&&  2 e^{-1/T} \cosh{\left(\frac{\Omega s}{T}\right)} \left[\tanh{\left ( \frac{\Omega s}{T}\right )}-m\right] \left(1-m s\right) \nonumber \ .
\end{eqnarray}

The effective temperature evolves according to the following configurational heat equation
\begin{eqnarray}
\label{dotchi}
&&c_0 \dot{\chi} =  2s\dot\gamma^{pl} \left(\chi_{\infty}-\chi \right) =\\
&& 2\epsilon_0 e^{-1/\chi} e^{-1/T} s \cosh{\left ( \frac{\Omega s}{T}\right )}\! \left[\tanh{\left( \frac{\Omega s}{T}\right)}\!-\!m\right](\chi_{\infty}\!-\!\chi) \nonumber \ ,
\end{eqnarray}
where the dimensionless plastic power $2\dot\gamma^{pl}s$ is the driving force and $c_0$ is a configurational heat capacity (dimensionless). As an increasing $\chi$ increases $\dot\gamma^{pl}$ in Eq. (\ref{Dpl}) through $e^{-1/\chi}$, Eq. (\ref{dotchi}) describes a ``flow induces flow'' mechanism (as long as $\chi\!<\!\chi_\infty$). $\chi_\infty$ is the steady-state value of $\chi$, reached at sufficiently long persistent plastic deformation, as observed in previous studies on unidirectional flows (see \cite{BLII-09} and references therein). It is taken here to be a material parameter independent of the dynamical variables in the problem. Note that ordinary temperature $T$ contributions to this heat equation are neglected as we focus here on low reservoir temperatures, and that $\chi$-diffusion is omitted as we focus on spatially homogeneous systems.

To close the set of equations we need an evolution equation for the stress $s$. To that aim, we need to make a kinematic decision about how to combine elastic and plastic deformation. We adopt here an additive decomposition of rates, $\dot\gamma\!=\! \dot\gamma^{el}+\dot\gamma^{pl}$, which is justified in our case as elastic strains remain small.
$\dot\gamma$ is the total strain rate, which in our case is externally controlled and takes the form $\dot\gamma\!=\!\omega\,\gamma_0\cos(\omega t)$ (recall that $\gamma\!=\!\gamma_0\sin(\omega t)$). The elastic strain rate $\dot\gamma^{el}$ is assumed to follow the linear elastic (in fact, linear hypo-elastic) relation $\dot{s}\!=\!2\mu\dot\gamma^{el}$, where $\mu$ is the shear modulus (dimensionless, in units of the yield stress $s_y$). Therefore, we obtain
\begin{eqnarray}
\label{dots}
\dot{s} = 2\mu\left[\omega \gamma_0 \cos{(\omega t)}- \dot\gamma^{pl}\right] \ .
\end{eqnarray}

Equations (\ref{dotm})-(\ref{dots}) (together with Eq. (\ref{Dpl})) constitute our elasto-viscoplasticity model. The power of the model, and also its limitations, stem from its relative simplicity that nevertheless generates a rather reach spectrum of physical behaviors. Here, like in \cite{Rycroft2012}, we treat the model as ``predictive'' in the sense that we do not aim at tuning the model's parameters to quantitatively fit a specific set of measurements or numerical simulations; rather we estimate the model's parameters from independent sources, and study its prediction for a new protocol (i.e. imposed variable-amplitude oscillatory shear).

We more or less use the same Bulk Metallic Glass parameters as in \cite{Rycroft2012}, except for the temperature $T$ which is taken to be rather small to mimic the athermal conditions used in experiments and numerical simulations, which at present offer the most comprehensive description of the phenomena of interest. In particular, the values of the dimensionless parameters are given in Table \ref{tab:parameters}. Note that the dimensionless value of the angular frequency $\omega$ corresponds to $1$sec$^{-1}$ in physical units, which means that the atomic vibrations time scale $\tau$ is set to $10^{-13}$sec.
\begin{table}[here]
 \centering
  \bgroup
\def\arraystretch{1.5}
 \begin{tabular}{|c|c|c|c|c|}
  \hline
 $~\mu = 43.53~$ &  $~\chi_{\infty}=0.0776~$ & $~\epsilon_0=0.1~$ & $~c_0=2~$ & $~\Omega = 0.7589~$ \\ \hline
 \end{tabular}
 \egroup
 \vspace{0.2cm}
  \bgroup
\def\arraystretch{1.5}
 \begin{tabular}{|c|c|}
  \hline
  $~T=0.0123~$ & $~\omega=10^{-13}~$\\ \hline
 \end{tabular}
 \egroup
  \caption{Material (top) and control (bottom) parameters used (dimensionless).}
 \label{tab:parameters}
\end{table}

The initial conditions are $\chi(0)\!=\!\chi_0$, $s(0)\!=\!0$, $m(0)\!=\!0$, which correspond to stress-free isotropic conditions. $\chi_0$ quantifies the initial state of disorder and reflects the preparation procedure of the amorphous material prior to the application of the oscillatory shear deformation. In most of the analysis below we use $\chi_0\!=\!0.0569\!<\!\chi_\infty$ (obtained through quenching from a relatively low temperature equilibrium state), but we also explore $\chi_0\!=\!0.0983\!>\!\chi_\infty$ (obtained through quenching from a relatively high temperature equilibrium state) situations (see Fig. \ref{fig3}).

\section{Theoretical analysis}
\label{sec:theory}

Our goal now is to study in detail, both analytically and numerically, the model discussed above. The analysis focuses on different amplitude regimes and the transitions between them, and addresses both steady- and non-steady-state characteristics of the oscillatory deformation.

\subsection{The linear response regime}
\label{subsec:LinearResponse}

We begin by considering the linear response regime of very small amplitudes $\gamma_0$, for which one can linearize Eqs. (\ref{dotm})-(\ref{dots}) to obtain
\begin{eqnarray}
&&\dot{m} \simeq 2\,e^{-1/T} \left( \frac{\Omega s}{T}-m\right) ,\\
&&\dot{\chi} \simeq {\C O}(\gamma_0^2) \ ,\\
&&\dot{s} \simeq 2\mu \left[ \omega \gamma_0 \cos{(\omega t)}\!-\! \epsilon_0 e^{-1/\chi_0}e^{-1/T} \left(\frac{\Omega s}{T}-m\right) \right] .
\end{eqnarray}

These linearized equations can be readily solved together with the initial conditions to yield
\begin{eqnarray}
\label{linear_solution}
\!\!\!\!s(t) &\simeq& 2\mu\,\gamma_0 \left(s_1\sin{(\omega t)} +s_2 \left[\cos{(\omega t)-e^{-\beta\,t}}\right] \right) \ ,\\
\!\!\!\!\chi(t) &\simeq& \chi_0 + {\C O}(\gamma_0^2) \ ,\\
\!\!\!\!m(t) &\simeq& \frac{2\mu\gamma_0\alpha\,\Omega}{T}\!\left(m_1 \sin{(\omega t)}\!-\!m_2 \left[\cos{(\omega t)\!-\!e^{-\beta\,t}}\right]\right) \ ,\nonumber\\
\end{eqnarray}
where
\begin{eqnarray}
\alpha &\equiv& 2\,e^{-1/T} \ , \quad \beta \equiv \alpha\left(1+\frac{\epsilon_0\Omega\mu e^{-1/\chi_0}}{T}\right) \ ,\\
s_1 &\equiv& 1 -\frac{\beta(\beta-\alpha)}{\beta^2+\omega^2} \ , \quad s_2 \equiv \frac{\omega (\beta-\alpha)}{\beta^2+\omega^2} \ , \\
m_1 &\equiv& \frac{\alpha\,s_1+\omega\,s_2}{\alpha^2+\omega^2} \ , \quad m_2 \equiv \frac{\omega\,s_1-\alpha\,s_2}{\alpha^2+\omega^2} \ .
\end{eqnarray}

For typical parameter values, such as those used in our calculations (cf. Table \ref{tab:parameters}), $s_2$ is extremely small and $s_1$ is very close to unity. Furthermore, the amplitude of $m(t)$ remains very small. Hence, in spite of the fact that the timescale to reach steady-state, $\beta^{-1}$, might be very large, for every practical purpose we can approximate the stress by a purely elastic response, $s(t)\!\simeq\!2\mu\,\gamma_0 \sin{(\omega t)}$. In addition, note that to linear order in $\gamma_0$, the structural state of disorder of the system does not evolve ($\chi(t)\!\simeq\!\chi_0$), though structural rearrangements continuously take place. Consequently, the system carries long time memory of its initial state characterized by $\chi_0$. These features remain valid beyond the linear response regime, as long as $\gamma_0$ is smaller than some critical amplitude $\gamma_c$, as will be discussed below.

The fact that dissipative rearrangements continuously occur is manifested in $m(t)$ being finite (yet extremely small). In the long time limit (when steady-state oscillatory conditions are reached), we can calculate the area $\C A$ under the hysteresis curve in steady-state
\begin{equation}
\label{A def}
{\C A} \equiv \lim_{t \to \infty} \oint s \dot{\gamma}\,dt \ ,
\end{equation}
which is proportional to the dissipated power per cycle, obtaining
\begin{equation}
\label{A}
{\C A} = 2\pi\mu s_2 \gamma_0^2+{\C O}(\gamma_0^4) \ .
\end{equation}
This shows that there is a finite (yet very small, as $s_2\!\ll\!1$) dissipation rate in the process. Moreover, as the dissipation rate $2s\dot\gamma^{pl}$ (recall that $s \dot\gamma^{el}$ does not contribute a finite power in a full cycle) drives the evolution of the structure in Eq. (\ref{dotchi}), it is clear that $\chi\!=\!\chi_0$ is not the true steady-state of the system (which is always $\chi\!=\!\chi_\infty$). Nevertheless, since the dissipation rate is extremely small, the evolution of structural disorder is enormously slow even compared to the relaxation time of $m(t)$, and hence the solution above can be regarded as a steady-state one for all practical purposes.

\subsection{The intermediate amplitudes regime: The onset of nonlinearity}
\label{usbsec:IntermediateAmplitudes}

Suppose now that $\gamma_0$ is increased beyond the linear response regime, but still remains not too large (in a sense that will be quantified later on). A necessary condition for linearity is $\Omega|s|/T\!\ll\!1$. Moreover, whenever the stress is such that $\Omega|s|/T$ is larger than unity, $\tanh{\left(\Omega s/T\right)}$ in Eq. (\ref{dotm}) essentially equals sign($s$). Under these conditions, one might naively expect that the steady-state amplitude of $m(t)$ equals $1$ (i.e. that $\dot{m}\!=\!0$ when $(\pm 1 - m)$ vanishes), similarly to the known behavior of the STZ model for unidirectional loading conditions. The situation, however, is somewhat more subtle. The point is that while it might be the case that during most of a cycle the stress is such that $\tanh{\left(\Omega s/T\right)}\!=\!\hbox{sign}(s)$, there might not be enough time for $|m(t)|$ to increase to unity (at a given oscillations frequency $\omega$). To estimate the typical stress amplitude $\tilde s$ for which the amplitude of $m(t)$ becomes $\C O(1)$, we ask what is the stress for which the product of the oscillations period $2\pi \omega^{-1}$ and the rate of growth of $m$ in Eq. (\ref{dotm}), $\case{1}{2}e^{-\frac{1-\Omega \tilde s}{T}}$ (recall that $\cosh{\left(\Omega s/T\right)}\!\simeq\!\case{1}{2}e^{\Omega s/T}$ for the stresses of interest), becomes $\C O(1)$, i.e.
\begin{equation}
\pi\,\omega^{-1}\,e^{-\frac{\displaystyle 1-\Omega \tilde s}{\displaystyle T}} \sim 1 \quad \Longrightarrow \quad \tilde s \sim \frac{1+T\ln\!\left(\frac{\omega}{\pi}\!\right)}{\Omega} \ .
\end{equation}

\begin{figure}[here]
\centering \epsfig{width=.5\textwidth,file=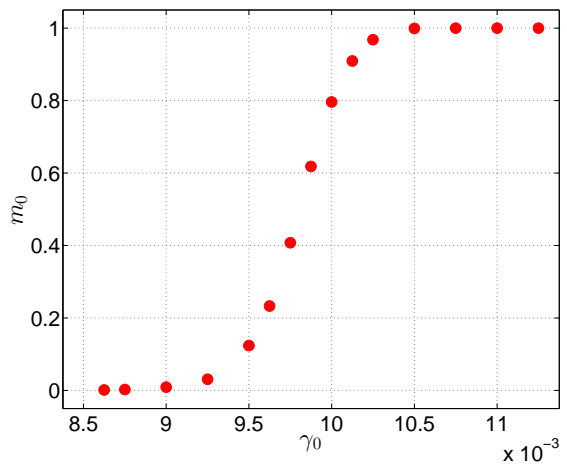}
\caption{The amplitude of $m(t)$ in the oscillatory steady-state, denoted by $m_0$, as a function of the strain amplitude $\gamma_0$, in the sub-yield regime.}
\label{fig1}
\end{figure}

An estimate of the typical strain amplitude $\tilde{\gamma}_0$ that corresponds to $\tilde s$ is simply obtained through $\tilde{\gamma}_0\!\simeq\!\frac{\tilde{s}}{2\mu}$, yielding $\tilde{\gamma}_0\!\simeq\!0.009$ for our parameters. Therefore, we expect the steady-state amplitude of $m(t)$ to remain much smaller than unity for $\gamma_0\!\ll\!\tilde{\gamma}_0$ and to be of order unity for $\gamma_0\!\gtrsim\!\tilde{\gamma}_0$. This is supported by the results presented in Fig. \ref{fig1}. When $\gamma_0$ becomes sufficiently larger than $\tilde{\gamma}_0$, we expect the amplitude of $m(t)$ to saturate at unity, i.e. $\dot{m}\!=\!0$ because $(\pm 1 - m)\!=\!0$, as in the unidirectional case. This is again supported by Fig. \ref{fig1} (see also Fig. \ref{fig2}). We remind the reader that $m\!=\!\pm 1$ physically means that all of the existing STZ are oriented in one direction.

In all of the solutions discussed in this section, plastic deformation remains very limited in magnitude and hence $\chi$ does not evolve substantially for an enormous number of loading cycles, remaining close to its initial value $\chi_0$. That is, the system in this regime carries memory of its history for very long times. Furthermore, the stress response is nearly purely elastic. This behavior breaks down at a critical amplitude $\gamma_c$, to be discussed next.

\subsection{The yielding transition and the large amplitudes regime}
\label{subsec:yield}

In the framework of the STZ model for unidirectional loading conditions, the evolution equation for $m(t)$, which reads $\dot{m}\!\sim\! [\hbox{sign}(s)-m][1-s\,m]$, exhibits an exchange of dynamic stability at $|s|\!=\!1$. when this happens, the fixed-point corresponding to $\dot{m}\!=\!0$ changes from $m\!=\!\hbox{sign}(s)$ (where the first brackets vanish) to $m\!=\!1/s$ (where the second brackets vanish). Since for the former $\dot\gamma^{pl}\!=\!0$, while for the latter $|\dot\gamma^{pl}|\!>\!0$, this was interpreted as a dynamic yielding transition and $|s|\!=\!1$ as a dynamic yield stress. A related behavior is observed in the oscillatory loading case.

\begin{figure}[here]
\centering \epsfig{width=.51\textwidth,file=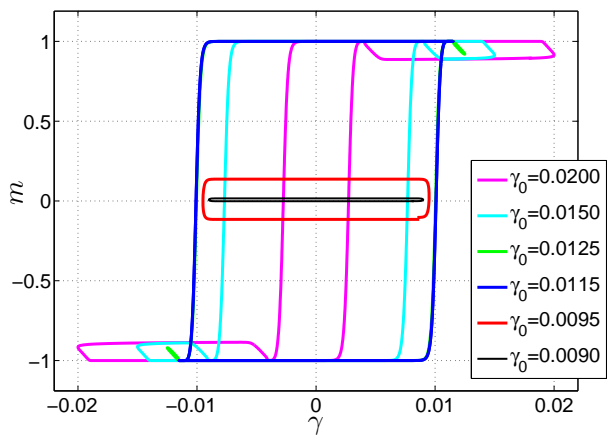}
\caption{$m(t)$ vs. $\gamma(t)$ over one cycle in steady-state for various $\gamma_0$'s. For $\gamma_0\!<\!\tilde\gamma$, the amplitude of $m(t)$ is very small (recall that $\tilde\gamma\!\simeq\!0.009$). For $\tilde\gamma\!<\!\gamma_0\!<\!\gamma_c$ the amplitude of $m(t)$ increases monotonically (cf. Fig. \ref{fig1}) until it saturates at unity (recall that $\gamma_c\!\simeq\!0.01155$). For $\gamma_0\!>\!\gamma_c$, finite time yielding periods, in which $|m|\!=\!1$ changes to $|m|\!<\!1$, are observed. The duration of these periods increases with increasing $\gamma_0$.}
\label{fig2}
\end{figure}

When the oscillations amplitude $\gamma_0$ is such that $|s|$ surpasses unity during a cycle, i.e. when $\gamma_0\!>\!\gamma_c\!\simeq\!(2\mu)^{-1}$ (as only very little plastic deformation takes place below the threshold, a purely elastic estimate is sensible), the system undergos a kind of yielding transition (for our parameters $\gamma_c\!\simeq\!0.01155$). That is, during part of the deformation cycle $m\!=\!1$ and then it switches to $m\!=\!1/s$ over another part, and the latter is accompanied by significant plastic flow. This is clearly observed in Fig. \ref{fig2}. The lifetime of the flowing state within a cycle increases with increasing $\gamma_0$, which is also demonstrated in Fig. \ref{fig2}. Note that this exchange of stability (bifurcation) is accompanied by rather strong variation of various physical quantities, as will be discussed below.

Since plastic deformation becomes significant for $\gamma_0\!>\!\gamma_c$, $\chi$ significantly evolves according to Eq. (\ref{dotchi}) until it reaches its true stable fixed-point, $\chi\!=\!\chi_\infty$. This means that the structure undergoes significant disordering (if $\chi_0\!<\!\chi_\infty$, the complementary case of $\chi_0\!>\!\chi_\infty$ will be discussed below as well) and the system loses memory of its initial state (characterized by $\chi_0$). In the following subsections, we aim at better characterizing both the transition at $\gamma_c$ and the behavior of the system in the $\gamma_0\!>\!\gamma_c$ regime.

\subsection{Transient dynamics: The timescale to approach steady-state below and above the critical amplitude}
\label{subsec:timescale}

Our goal here is to characterize the transient dynamics both below and above the critical amplitude $\gamma_c$. In particular, we focus on the typical time it takes the system to reach steady-state in both the post- and sub-yield regimes.

\subsubsection{Approach to steady-state in the post-yield regime}
\label{subsubsec:t post-yield}

\begin{figure}[here]
\centering \epsfig{width=.52\textwidth,file=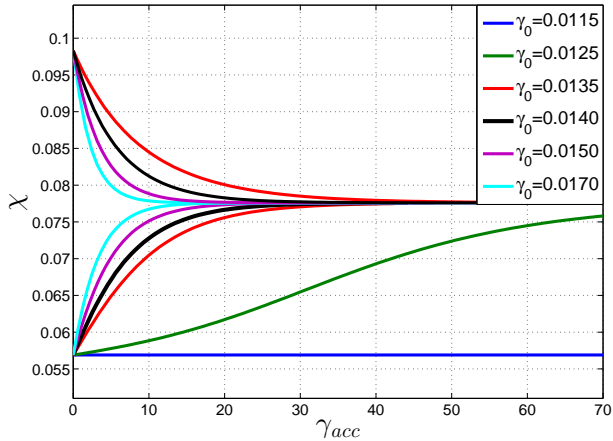}
\caption{$\chi$ vs. the accumulated strain $\gamma_{acc}\!\equiv\!\int_0^t |\dot\gamma|dt$ for various $\gamma_0\!\gtrsim\!\gamma_c$, for both $\chi_0\!=\!0.0569\!=\!\chi_\infty\!-\!0.0207\!<\!\chi_\infty$ and $\chi_0\!=\!0.0983\!=\!\chi_\infty\!+\!0.0207\!>\!\chi_\infty$. For sufficiently large $\gamma_0$'s, the fixed-point $\chi_{\infty}=0.0776$ is reached within the range of $\gamma_{acc}$ shown. For smaller $\gamma_0$'s, closer to $\gamma_c$, the fixed-point is not yet reached.}
\label{fig3}
\end{figure}

In the post-yield regime, i.e. for $\gamma_0\!\ge\!\gamma_c$, the system approaches a steady-state on a timescale that is determined by Eq. (\ref{dotchi}). In Fig. \ref{fig3} we show the time evolution (quantified here, as in \cite{Fiocco2013}, by the accumulated strain defined as $\gamma_{acc}\!\equiv\!\int_0^t |\dot\gamma|dt$) of $\chi$ for various values of $\gamma_0\!\ge\!\gamma_c$. Two initial conditions are used, one such that $\chi_0\!<\!\chi_\infty$ (obtained through quenching from a relatively low-temperature equilibrium state) and the other $\chi_0\!>\!\chi_\infty$ (obtained through quenching from a relatively high-temperature equilibrium state). The first observation, which is straightforward in light of Eq. (\ref{dotchi}), is that the system approaches a steady-state of disorder characterized by $\chi_\infty$, irrespective of both the initial condition $\chi_0$ and the oscillations amplitude $\gamma_0$. In the analysis below, we focus on the $\chi_0\!<\!\chi_\infty$ case. The second observation is that there is huge variation in the relaxation timescale (accumulated strain) to steady-state as a function of $\gamma_0$. In Fig. \ref{fig3}, we integrated the model equations up to an accumulated strain level $\gamma_{acc}$ that allowed $\chi$ to reach its steady-state for sufficiently large $\gamma_0$, but for $\gamma_0\!\approx\!\gamma_c$, $\chi$ barely evolves (see lowest curve).

\begin{figure}[here]
\centering \epsfig{width=.52\textwidth,file=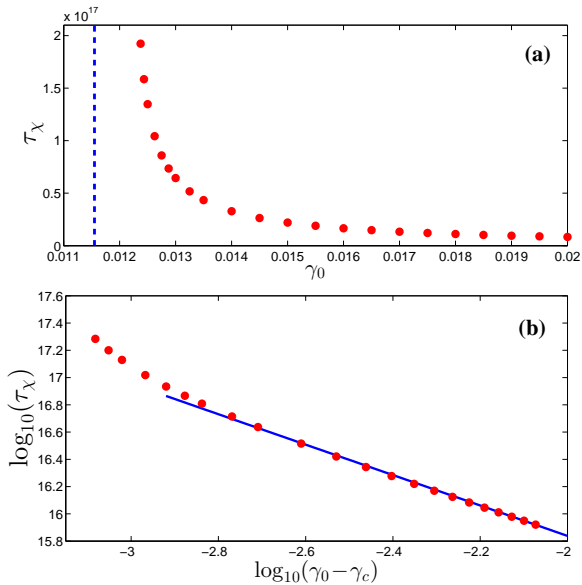}
\caption{(a) The relaxation time to steady-state $\tau_\chi$ as a function of $\gamma_0$. The dashed vertical line corresponds to $\gamma_0\!=\!\gamma_c$.  (b) The same data as in (a), but on log-log scale. The solid line has a slope of $-1.1$, which is consistent with the theoretical prediction in Eq. (\ref{eq:tauchi prediction}), $\tau_\chi\!\sim\!\left(\gamma_0\!-\!\gamma_c\right)^{-1}$.}
\label{fig4}
\end{figure}

To quantify the latter behavior we plot in Fig. \ref{fig4} the relaxation time of $\chi$, $\tau_\chi$ (defined as the time it takes $\chi$ to approach $\chi_\infty$ to $99\%$ of $\chi_\infty\!-\!\chi_0$), as a function of $\gamma_0$. Not very close to $\gamma_c$, we expect $\tau_\chi$ to drop off approximately as $\left(\gamma_0\!-\!\gamma_c\right)^{-1}$ (to be explained below), which is in reasonable agreement with the numerical data shown in Fig. \ref{fig4}b. As $\gamma_c$ is approached (from above), $\tau_\chi$ appears to exhibit a stronger variation with $\gamma_0$ than $\left(\gamma_0\!-\!\gamma_c\right)^{-1}$.

It is important to note that it is numerically difficult to actually approach $\gamma_c$ (recall that for the used parameters we have $\gamma_c\!\simeq\!0.01155$), where $\tau_\chi$ increases dramatically. Furthermore, note that while in the model $\tau_\chi$ does not strictly diverge at $\gamma_c$, in practical terms the growth of $\tau_\chi$ as $\gamma_0\!\to\!\gamma_c^+$ is so strong that it might appear as a singularity.

\subsubsection{Approach to steady-state in the sub-yield regime}
\label{subsubsec:t sub-yield}

In the sub-yield regime, i.e. for $\gamma_0\!<\!\gamma_c$, the system undergoes very little plastic deformation and hence to a very good approximation we have $\chi(t)\!\simeq\chi_0$ and $s(t)\!\simeq\!2\,\mu\,\gamma_0 \sin{(\omega t)}$ for extremely long times. Therefore, we treat $\chi(t)$ and $s(t)$ as fixed at these quasi-stationary states and consequently the approach of the system to a quasi-steady-state in the sub-yield regime is wholly determined by the evolution of $m(t)$ according to Eq. (\ref{dotm}).

To gain physical (and mathematical) insight into the dynamics of the orientational order parameter $m(t)$ in the sub-yield regime we consider first the simpler case of unidirectional shearing, for which Eq. (\ref{dotm}) reads
\begin{eqnarray}
\label{eq:dotm unidirectional}
\dot{m} = {\cal C}(T, s) \left(1-m\right) \left(1-s\,m\right) \ ,
\end{eqnarray}
where we defined the stress-biased thermal activation rate factor as ${\cal C}(T, s)\!\equiv\!2\,e^{-1/T}\cosh{\left(\Omega s/T\right)}$ and replaced $\tanh{\left(\Omega s/T\right )}$ with unity, as stresses such that $s\!\gg\!T/\Omega$ are of interest here. As long as $s\!<\!1$, i.e. in the sub-yield regime, the stable fixed-point of $m$ is $m\!=\!1$. In analogy with the oscillatory case (where $s(t)\!\simeq\!2\,\mu\,\gamma_0 \sin{(\omega t)}$ in the sub-yield regime), we are interested in the evolution of $m$, from $m(t\!=\!0)\!=\!0$ to $m\!\to\!1$, when the stress takes the form $s(t\!>\!0)\!=\!s_0\!<\!1$; that is, when the stress is abruptly ramped at $t\!=\!0$ to a constant smaller than the yield stress.

Under these conditions, the solution of Eq. (\ref{eq:dotm unidirectional}) can be readily obtained analytically, and for our purposes here it is best presented in the form
\begin{eqnarray}
\label{eq:dotm unidirectional sol}
\!\!\!\!\!\!\!\!\!t(m,s_0)\!=\! \frac{\ln\!\left(\frac{\displaystyle 1-s_0\,m}{\displaystyle 1-m}\right)}{{\cal C}(T, s_0)\left(1-s_0\right)}\!\equiv\! \left[{\cal C}(T, s_0)\right]^{-1}\!\tau_m(m,s_0).
\end{eqnarray}
We observe that the time to reach a certain value of $m$ under a given stress $s_0$, namely $t(m,s_0)$, is a product of an STZ-related time $\tau_m(m,s_0)$ and an inverse stress-dependent thermal activation rate $\left[{\cal C}(T,s_0)\right]^{-1}$.
The latter decreases exponentially with increasing stress $s_0$, hence strongly reducing $t(m,s_0)$ as $s_0$ increases. As it is an overall pre-factor in Eq. (\ref{eq:dotm unidirectional}), which strongly ``screens'' the STZ contribution associated with the $m$-dynamics, $\tau_m(m,s_0)$, we would like first to focus on the latter.

\begin{figure}[here]
\centering \epsfig{width=.52\textwidth,file=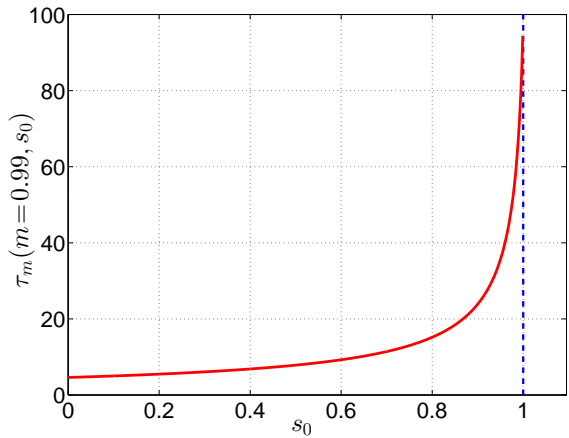}
\caption{The normalized time $\tau_m(m\!=\!0.99,s_0)$ (where the thermal activation rate factor is suppressed, cf. Eq. \ref{eq:dotm unidirectional sol}) it takes an initially isotropic system, $m(t\!=\!0)$, to reach $m\!=\!0.99$ as a function of $s_0$, the magnitude of an applied {\em unidirectional} (i.e. {\em not oscillatory}) shear stress in the sub-yield regime, $s_0\!<\!1$.}
\label{fig5}
\end{figure}

In Fig. \ref{fig5} we plot $\tau_m(m,s_0)$ of Eq. (\ref{eq:dotm unidirectional sol}) for $m\!=\!0.99$ (i.e. within 1\% from the fixed-point at $m\!=\!1$) as a function of the stress $s_0$. We observe that as the yielding transition is approached from below, $s_0\!\to\!1^-$, $\tau_m(m\!=\!0.99,s_0)$ increases significantly. This clearly shows that the STZ-mechanism embodied in the $m$-dynamics naturally produces slowing-down associated with the exchange of stability (bifurcation) taking place at the yielding transition, $s_0\!=\!1$. When the overall relaxation time $t(m,s_0)$ in Eq. (\ref{eq:dotm unidirectional sol}) is considered, i.e. when $\tau_m(m,s_0)$ is multiplied by the exponentially decreasing term $\left[{\cal C}(T,s_0)\right]^{-1}$, there will be a competition between the two contributions and depending on parameters, $t(m,s_0)$ may decrease with increasing $s_0$ over some range.

The oscillatory shear case is more mathematically involved as compared to the unidirectional one, yet we believe that the picture described above remains qualitatively similar. For the set of parameters used in our calculations, and within a range of shear amplitudes $\gamma_0$'s over which we managed to perform numerical calculations, the exponential rate factor dominated the relaxation time, which consequently decreases with increasing $\gamma_0$. Through some asymptotic analysis in the limit $\gamma_0 \!\to\!\gamma_c^-$ (not discussed here), we found evidence for an increase in the relaxation time of $m$ (similarly to the behavior in Fig. \ref{fig5}), due to the slowing-down associated with the exchange of stability at $\gamma_0\!=\!\gamma_c$, even in the presence of the rate factor. Whether this behavior might be somehow related to the experimental and simulational observations of an increasing relaxation time to steady-state when $\gamma_0 \!\to\!\gamma_c^-$, will be discussed below in sect. \ref{subsec:comparison_subyield}.

\begin{figure}[here]
\centering \epsfig{width=.5\textwidth,file=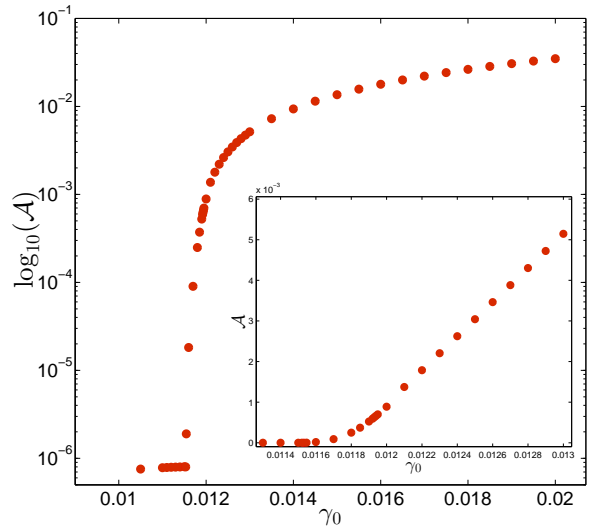}
\caption{(main) The logarithm of the dissipation per cycle in steady-state $\log_{10}({\C A})$ vs. $\gamma_0$, both in the sub-yield regime close to the transition ($\gamma_0\!\lesssim\!\gamma_c$) and in the post-yield regime ($\gamma_0\!>\!\gamma_c$). (inset) The same as the main panel, but with $\C A$ being plotted in linear scale. In the inset it is observed that while the transition is sharp, it is still continuous and differentiable. Above the transition, but not too close to it, $\C A$ approximately follows $\C A \!\sim\! \gamma_0 \!-\! \gamma_c$.}
\label{fig6}
\end{figure}

\subsection{Energy dissipation in steady-state}
\label{subsec:hysteresis}

In order to better quantify the continuous -- yet quite sharp -- transition near $\gamma_c$, we focus here on the dissipation per cycle in steady-state, quantified by the area under the hysteresis curve ${\cal A}$ defined in Eq. (\ref{A def}), as a function of $\gamma_0$. This quantity is of basic interest in general, and especially in the thermodynamic STZ model, where it is assumed to directly drive the evolution of structural disorder, i.e. of $\chi$ in Eq. (\ref{dotchi}).

\begin{figure}[here]
\centering \epsfig{width=.52\textwidth,file=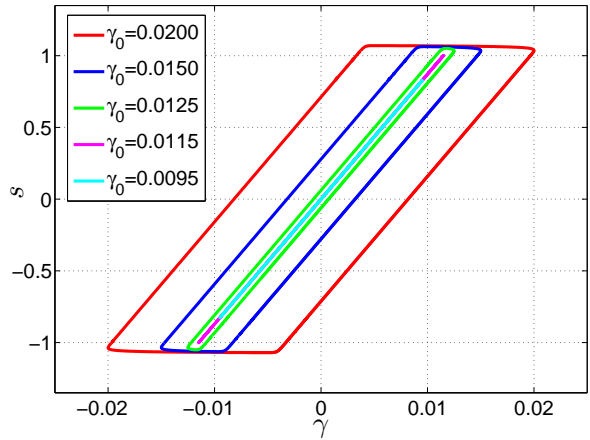}
\caption{Stress-strain curves for various $\gamma_0$'s, ranging from below $\gamma_c$ ($\gamma_0\!=\!0.0095$) to well above it ($\gamma_0\!=\!0.02$).}
\label{fig7}
\end{figure}

In Fig. \ref{fig6} we plot $\C A$ vs. $\gamma_0$. First, we observe that $\C A$ changes very sharply, by several orders of magnitude, near $\gamma_c$. Second, sufficiently above the transition, we have approximately $\C A\!\sim\! \gamma_0\!-\!\gamma_c$, clearly observed in the inset. The physical reason for this scaling behavior is that above the yielding transition, the stress amplitude remains roughly constant at a value slightly above the yield stress, while the strain amplitude grows linearly with $\gamma_0$. This is clearly seen in the stress-strain curves shown in Fig. \ref{fig7}. While well below the yielding transition, both the stress and the strain amplitudes significantly increase with $\gamma_0$, above it -- but not very close to it -- only the strain amplitude does. In the latter regime, the dissipation rate $2s\dot\gamma^{pl}$ -- which controls the evolution of $\chi$ in Eq. (\ref{dotchi}) -- can be roughly estimated by its steady-state value (of course when it strictly reaches steady-state, so does $\chi$). Consequently, Eq. (\ref{dotchi}) can be roughly approximated by $c_0 \dot{\chi} \!\simeq\! 2{\C A} \left(\chi_{\infty}-\chi \right)$, which suggests that in this regime the relaxation time $\tau_\chi$ approximately scales as
\begin{equation}
\label{eq:tauchi prediction}
\tau_\chi \sim {\C A}^{-1} \sim \left(\gamma_0- \gamma_c\right)^{-1} \ .
\end{equation}

This prediction is supported by the numerical results shown in Fig. \ref{fig4}b, where a power-law with an exponent $-1.1$ is observed. The possible relation of this theoretical result to experimental observations will be discussed in Sect. \ref{subsec:comparison_postyield}. With this we conclude the analysis of the model. Next, we aim at comparing the qualitative predictions of the model to the salient features emerging from recent computer simulations and experiments.

\section{Relation to simulations/experiments}
\label{sec:exp}

Our goal here is to discuss the model's predictions in relation to the main observations of relevant simulations and experiments \cite{Hebraud1997, Petekidis2002, Lacks2004, Lundberg2008, Slotterback2012, Ren2013, Priezjev2013, Keim2013a, Fiocco2013, Regev2013, Keim2014, Priezjev2014, Nagamanasa2014}. We discuss separately the sub- and post-yield regimes.

\subsection{The sub-yield regime}
\label{subsec:comparison_subyield}

As was discussed earlier, for strain amplitudes $\gamma_0$ smaller than $\gamma_c$, the amorphous system settles into dissipative limit cycles in which plastic rearrangements take place back-and-forth such that particles return to their positions after an integer number of loading cycles (at finite temperatures the system can jump between different limit cycles \cite{Regev2013}). The microscopic states visited by the system in this regime are correlated with the initial state, i.e. the system carries memory of its history. This behavior is consistent with the phenomenology emerging from the solutions of the STZ model in this regime, where the anisotropy order parameter $m$ settles into an oscillatory steady-state of zero average over a cycle (cf. Fig. \ref{fig2}), corresponding to plastic rearrangements that are reversed when the loading changes its sign.

This agreement lends serious support to the conceptual picture embodied in the STZ model in which flow defects in amorphous systems are characterized by internal states between which they can make transitions in response to external forcing or thermal fluctuations. That is, the existence of a zone should be distinguished from its internal states, an idea that is not always incorporated in existing theoretical efforts. In this regime, after a transient, zones are not being created and/or annihilated anymore (i.e. they retain their identity), rather they only change their internal state. This is directly manifested in the vanishing diffusion coefficient in the sub-yield regime, as observed in \cite{Fiocco2013}.

While STZ transitions are definitely dissipative, the dissipation per cycle in the sub-yield regime is extremely small. Since it is the dissipation that drives the evolution of the structure quantified in our framework by $\chi$ in Eq. (\ref{dotchi}), the system will remain close to its initial condition $\chi\!\simeq\!\chi_0$ for extremely long times. In this sense, the model agrees with the observation that the system carries long time memory of its initial state. Note that strictly speaking, since the dissipation is non-vanishing, Eq. (\ref{dotchi}) will eventually lead to $\chi\!=\!\chi_\infty\!\ne\!\chi_0$ as $t\!\to\!\infty$. This might be a rather academic issue as the timescale to reach this true steady-state is enormously large.

Simulational and experimental results in the low amplitudes regime also indicate that different limit cycles are reached as $\gamma_0$ is increased, and more importantly, that the time it takes to reach these limit cycles increases significantly as $\gamma_0$ increases. Some authors suggested that this slowing-down is characterized by a power-law divergence as $\gamma_0\!\to\!\gamma_c^-$ \cite{Regev2013, Nagamanasa2014}. Indeed, the STZ model predicts different steady-states corresponding to different $\gamma_0$'s in this regime, as shown in Fig. \ref{fig1}. The issue of the slowing-down as $\gamma_c$ is approached, which is one of the most interesting simulational and experimental observations, is more subtle.

The unidirectional results discussed in relation to Eq. (\ref{eq:dotm unidirectional sol}) and Fig. \ref{fig5} clearly show that the STZ-dynamics feature a slowing-down mechanism associated with the exchange of stability occurring at $\gamma_c$. As mentioned in Sect. \ref{subsubsec:t sub-yield}, we found evidence for a similar slowing-down behavior, dominated by a $(\gamma_c\!-\!\gamma_0)^{-1}$ variation of the relaxation time, also in the oscillatory case (though it was difficult to detect it numerically as the thermal activation rate factor, which may or may not be physically relevant to the computer simulations and experiments discussed here, exponentially attenuates the relaxation time).

While we find this slowing-down mechanism interesting, it is not entirely clear to us whether it is directly responsible for the experimentally and simulationally observed slowing-down. The point is that we know that a physical ingredient, that is relevant at least to the linear response regime, is missing from the ``plain vanilla'' STZ model analyzed here. That is, this model incorporates only a single activation energy barrier, rather that a distribution of such barriers, implying that the linear response modulus $G^*(\omega)$ will exhibit a single timescale Maxwell behavior as a function of frequency, which is manifestly inconsistent with the broad spectrum of relaxation times observed in glassy systems \cite{LinearResponse_PRE_2011, LinearResponse_PRL_2011}. The latter is directly related to the complexity of the potential energy landscape, which in turn might be linked to the increasing relaxation time with increasing $\gamma_0$. In particular, it is conceivable that as $\gamma_0$ increases, the system has more energy to explore regions of the potential energy landscape with more minima, and consequently it might take more time to settle into a limit cycle.

Finally, we note that in \cite{Nagamanasa2014} it was suggested that the growing relaxation time is associated with a growing correlation lengthscale due to interaction between different rearrangements, a feature that is absent from the mean-field STZ model. It also worth noting that while the STZ slowing-down mechanism is related to the dynamics of the orientational order parameter $m$, measurements of a growing relaxation timescale were based either on the evolution of the energy \cite{Regev2013, Fiocco2013} or on the fraction of particles that do not return to their initial positions at the end of a strain cycle \cite{Nagamanasa2014}. We have not discussed possible relations between these different quantities, and have not addressed the question of whether different physical quantities can exhibit different relaxation times. It is worth noting here that $m$, or similar anisotropy-related coarse-grained order parameters (internal variables), are not yet measured directly in simulations and experiments. Defining and measuring such an internal variable remains a major open challenge in the field.

\subsection{The yielding transition and the post-yield regime}
\label{subsec:comparison_postyield}

Simulations and experiments clearly demonstrate a qualitative change in the response of the deforming system when the amplitude $\gamma_0$ surpasses a threshold value $\gamma_c$. That is, the system undergoes a yielding transition. When this happens, particles start to diffuse, the system loses memory of its initial conditions and significant dissipation takes place. Similarly to previous unidirectional shear analyses based on the STZ model, such a yielding transition spontaneously emerges also in the oscillatory shear case, cf. Sect. \ref{subsec:yield}. It is important to stress here again that this yielding transition emerges due to a dynamic exchange of stability (bifurcation) in the model's equations, and not due to a threshold that is introduced in an ad hoc manner.

While this is a neat scenario of a yielding transition, we believe that better understanding the slowing-down of the dynamics in the sub-yield regime (discussed above in Sect. \ref{subsec:comparison_subyield}), and its relation to the yielding transition, can shed light on whether the STZ yielding picture is physically sensible or not. One can imagine, for example, that yielding requires going beyond a mean-field description and/or is intimately related to changes in the distribution of activation barriers. This is an important direction of future investigation.

We now aim at discussing the model's predictions in the post-yield regime in relation to several salient features observed in recent simulations and experiments:\\

{\bf $\bullet$ The existence of a unique stochastic structural steady-state independent of $\gamma_0$} --- The STZ model, following Eq. (\ref{dotchi}), predicts that for sufficiently long and slow oscillatory shear deformation, the system approaches a unique stochastic structural state of disorder characterized by $\chi_\infty$, independently of $\gamma_0$ and the initial conditions. This is clearly observed in Fig. \ref{fig3}. This issue has been partially addressed in the athermal quasi-static atomistic simulations of \cite{Fiocco2013}. The results presented in \cite{Fiocco2013}, cf. Fig. 1a therein, might give the impression that the steady-state structural state is independent of initial conditions (e.g. whether $\chi_0\!<\!\chi_\infty$ or $\chi_0\!>\!\chi_\infty$, as predicted by the thermodynamic STZ model), but that it depends on the amplitude $\gamma_0$. This is not obvious, however, since the quantity presented there is the potential energy at zero strain, $E(\gamma\!=\!0)$. As the latter includes also an elastic contribution that is expected to scale as $[s(\gamma\!=\!0)]^2/\mu$, this is not a purely structural measure. In fact, as the elastic energy at $\gamma\!=\!0$ increases with increasing $\gamma_0$, we expect $E(\gamma\!=\!0)$ to increase with increasing $\gamma_0$, which is indeed observed in the simulations \cite{Fiocco2013}. Another possible source of bias in addressing this important question is that many of the simulational works are performed under NVT conditions in which the pressure may vary, yet again contributing a variable elastic energy (this time dilatational, not distortional) to the total potential energy. This point, i.e. determining whether the stochastic structural steady-state depends on the amplitude $\gamma_0$ or not, should be clarified by simulations and experiments, preferably by measuring purely structural quantities (e.g. the shear modulus) instead of the potential energy.\\

{\bf $\bullet$ The relaxation time to steady-state} --- The relaxation time to steady-state in the post-yield regime, $\tau_\chi$, is shown in Fig. \ref{fig4}. Not very close to $\gamma_c$, it falls off with increasing $\gamma_0$ as $\tau_\chi\!\sim\!\left(\gamma_0\!-\!\gamma_c\right)^{-1.1}$, quite similarly to the theoretical estimation in Eq. (\ref{eq:tauchi prediction}). Interestingly, a power-law variation with an exponent of $-1.1\pm0.3$ (not very close to the transition amplitude) has been very recently reported in experiments on a colloidal glass \cite{Nagamanasa2014}. If this agreement is not superficial, then the thermodynamic STZ model offers an intuitive and physically transparent explanation for the $\gamma_0$-dependence of the relaxation time in the post-yield regime, not very close to the transition. Fig. \ref{fig4}b indicates that as $\gamma_0\!\to\!\gamma_c^+$, the relaxation time increases even stronger with decreasing $\gamma_0$, apparently corresponding to the power-law being augmented by exponential growth. The stronger increase is understood since the dissipation per cycle, which controls the evolution of $\chi$, can no longer be reasonably approximated by $2\C A$. Extracting the data of Fig. 1b in \cite{Fiocco2013}, we indeed found a stronger than $\tau_\chi\!\sim\!\left(\gamma_0\!-\!\gamma_c\right)^{-1.1}$ increase in the relaxation time, as predicted by our STZ analysis. It would be fair to note that while the experimental and simulational results in this respect are suggestive, they typically span too limited a range of amplitudes to decisively determine power-laws and divergent behaviors.\\

{\bf $\bullet$ The dissipation per cycle in steady-state} --- The dissipation per cycle in steady-state, $\C A$, is a macroscopic quantity that can be useful in characterizing the yielding transition. As observed in Fig. \ref{fig6}, $\C A$ is extremely small in the sub-yield regime and it increases substantially above the yielding transition. The model predicts a smooth, yet very strong, increase in $\C A$ near $\gamma_c$ and $\C A \!\sim\! \gamma_0\!-\!\gamma_c$ away from the transition. This is consistent with the data presented in Fig. 3b of \cite{Fiocco2013} and is not surprising; it simply results from the fact that while the amplitude of the stress almost saturates with increasing $\gamma_0$, the amplitude of the strain grows linearly with it in this regime (cf. Fig. \ref{fig7} here and the inset of Fig. 3b in \cite{Fiocco2013}). It would be interesting to see whether more detailed measurements (either in simulations or in experiments) of $\C A$ near $\gamma_c$ can teach us more about the yielding transition and the status of the theoretical picture demonstrated in Fig. \ref{fig6}.

\section{Concluding remarks}
\label{sec:sum}

In this paper we presented a rather extensive analysis of the ``plain vanilla'' thermodynamic STZ model for a variable-amplitude oscillatory shear protocol, and discussed the results in light of very recent experimental and simulational observations. The comparison highlights both the success of the model and its possible limitations.

The existence of dissipative limit cycles in the low-amplitude regime, in which back-and-forth localized particle rearrangements periodically bring the system to the same state and in which the system retains memory of its initial state (i.e. remains closely correlated with the initial condition), clearly indicates that STZ can retain their identity while changing their internal state (``polarization'') in response to stress. This lends direct support to the idea that one should distinguish between STZ creation/annihilation and transitions between the internal states of an STZ. While this is a basic ingredient in the thermodynamic STZ model, other models do not make this important distinction (e.g. an STZ is assumed to be annihilated after a transition and a new, uncorrelated, STZ is assumed to be created instead). As the thermodynamic STZ model assumes that structural evolution is driven by dissipation, and as the dissipation in the low-amplitude regime is very small (yet finite, as it does involve STZ transitions), memory of the initial state is retained.

The thermodynamic STZ model exhibits a yielding transition at a critical amplitude $\gamma_c$, which is associated with an exchange of stability (bifurcation) in the dynamic equation for the orientational order parameter $m$. The dissipation per cycle in steady-state increases as ${\C A}\!\sim\!\gamma_0-\gamma_c$ above the transition, as observed in simulations \cite{Fiocco2013}. The model predicts a strong increase in the relaxation time to steady-state as $\gamma_c$ is approached from above, as observed experimentally and simulationally \cite{Fiocco2013, Regev2013, Nagamanasa2014}. The idea that dissipation drives structural evolution suggests that not too close to $\gamma_c$ the relaxation time approximately behaves as $\tau_\chi \!\sim\! (\gamma_0-\gamma_c)^{-1}$, which might be consistent with some experimental data \cite{Nagamanasa2014}.

The model predicts that above the transition amplitude $\gamma_c$, systems quenched from very different equilibrium liquid states at possibly different rates, all approach a stochastic oscillatory steady-state independent of the initial state. This agrees with simulational results \cite{Fiocco2013}. The model also predicts that this steady-state is independent of the oscillations amplitude $\gamma_0\!>\!\gamma_c$. We believe that currently available experimental and simulational data are still inconclusive in relation to this prediction.

One of the most interesting experimental and simulational results is the observation of a growing relaxation time as the threshold amplitude approached from below, $\gamma_0\!\to\!\gamma_c^-$. In our view, understanding this behavior, and its possible relation to the yielding transition itself, is important. We have shown that the STZ model features a slowing-down mechanism associated with the exchange of stability in the dynamics of $m$. Yet, it is unclear to us whether it is directly related to the experimental and simulational observations. In particular, one may wonder whether a broad distribution of activation barriers \cite{SOLLICH-97, LinearResponse_PRL_2011, LinearResponse_PRE_2011}) and STZ-STZ interactions do not have bearings on the slowing-down issue. This should be addressed in a future investigation.

We believe that the present analysis of the thermodynamic STZ model offers a sensible continuum-level framework to address the variable-amplitude oscillatory shear response of amorphous materials. As such, it may serve as a starting point for additional theoretical developments. A fruitful direction may be to find experimental and simulational procedures to measure $m$, or a similar anisotropy-related internal variable, and the effective temperature $\chi$.\\

{\em Acknowledgements}. We thank S. Sastry, D. Fiocco, G. Foffi, E. Lerner and I. Regev for valuable discussions.
We acknowledge support from the Minerva Foundation with funding from the Federal German Ministry for Education and Research, the Israel Science Foundation (Grant No. 712/12), the Harold Perlman Family Foundation and the William Z. and Eda Bess Novick Young Scientist Fund.

\end{document}